\documentclass[iop]{emulateapj}
\usepackage{hyperref}
\usepackage{amsmath}
\usepackage{amssymb}
\usepackage{graphicx}

\newcommand{\kms}{km\,s$^{-1}$}
\newcommand{\Teff}{$T_{\rm eff}$}
\newcommand{\logg}{$\log g$}
\newcommand{\micro}{$\nu_{\rm micr}$}

\newcommand{\TucA}{TucII-006}
\newcommand{\TucB}{TucII-011}
\newcommand{\TucC}{TucII-033}
\newcommand{\TucD}{TucII-052}

\begin{document}
\shorttitle{Tucana II Chemical Abundances}
\title{Chemical Diversity in the Ultra-faint Dwarf Galaxy Tucana~II\altaffilmark{*}}

\author{Alexander P. Ji\altaffilmark{1,2}, 
		Anna Frebel\altaffilmark{1,2}, 
        Rana Ezzeddine\altaffilmark{2,1}, and 
        Andrew R. Casey\altaffilmark{3}}
\altaffiltext{*}{This paper includes data gathered with the 6.5 m Magellan Telescopes
located at Las Campanas Observatory, Chile.}
\altaffiltext{1}{
  Department of Physics and Kavli Institute for Astrophysics and Space
  Research, Massachusetts Institute of Technology, Cambridge, MA
  02139, USA; \texttt{alexji@mit.edu}}
\altaffiltext{2}{
  Joint Institute for Nuclear Astrophysics - Center for Evolution of the Elements, East Lansing, MI 48824}
\altaffiltext{3}{
  Institute of Astronomy, University of Cambridge, Madingley Road, Cambridge, CB3 0HA, United Kingdom}

\begin{abstract}
We present the first detailed chemical abundance study of the ultra-faint dwarf galaxy Tucana~II based on high-resolution Magellan/MIKE spectra of four red giant stars.
The metallicity of these stars ranges from $\mbox{[Fe/H]} = -3.2$ to $-2.6$, and all stars are low in neutron-capture abundances ([Sr/Fe] and [Ba/Fe] $< -1$).
However, a number of anomalous chemical signatures are present.
Three stars are carbon-enhanced, including the most metal-rich star. This star ($\mbox{[Fe/H]}=-2.6$)  shows [Na,$\alpha$,Sc/Fe] $< 0$, suggesting an extended star formation history with contributions from AGB stars and Type~Ia supernovae.
The other carbon-enhanced stars have $\mbox{[Fe/H]} < -3$ and may be consistent with enrichment by faint supernovae, if such supernovae can produce neutron-capture elements.
A fourth star with $\mbox{[Fe/H]} = -3$ is carbon-normal, and exhibits distinct light element abundance ratios from the carbon-enhanced stars.
The carbon-normal star implies that at least two distinct nucleosynthesis sources, both possibly associated with Population~III stars, contributed to the early chemical enrichment of this galaxy.
Despite its very low luminosity, Tucana~II shows a diversity of chemical signatures that preclude it from being a simple ``one-shot'' first galaxy, but still provide a window to star and galaxy formation in the early universe.
\end{abstract}


\section{Introduction}\label{s:intro}
Ultra-faint dwarf galaxies (UFDs) are old, metal-poor galaxies with large mass-to-light ratios \citep{Simon07,Brown14}.
These galaxies are ${>}30$\,kpc away, but detailed chemical abundances can be derived for the brightest stars in UFDs through high-resolution spectroscopy on 10\,m class telescopes.
The abundances of these metal-poor stars likely trace the nucleosynthetic output of the first Population~III (Pop~III) stars that enriched their host galaxy. 
Since UFDs have relatively simple star formation histories, they are a particularly powerful probe for dwarf galaxy archaeology, as all their stars formed from the same galactic environment \citep[e.g.,][]{Frebel12,Karlsson13,Ji15}. 
This provides valuable constraints on the nature and site of the first nucleosynthesis events that cannot be derived for field stars from the chemical signatures alone \citep[e.g.,][]{Ji16b}.

High-resolution spectroscopy has led to elemental abundance measurements of stars in ten different UFDs. 
The overarching message is that, in most respects, stars in UFDs are chemically similar to ordinary metal-poor halo stars. 
Considering the population of ten UFDs, the lowest metallicity stars tend to be carbon-enhanced, a likely signature of the first stars \citep[e.g.,][]{Cooke14, Placco14, Ji15, Yoon16}.
Most UFDs show evidence for somewhat sustained star formation and chemical evolution, with [$\alpha$/Fe] ratios that decline over the range $\mbox{[Fe/H]} =-3$ to $-2$, with the notable exception of Segue~1 \citep{Vargas13,Frebel14}\footnote{$\mbox{[X/Y]} = \log_{10}(N_X/N_Y) - \log_{10}(N_X/N_Y)_\odot$ for elements X,Y}.
The overall duration of star formation is expected to be very short \citep{Brown14,Webster15}, and these galaxies appear to completely lack stars with $\mbox{[Fe/H]} \gtrsim -1.5$.
However, the heavy element abundances of UFD stars differ significantly from those of halo stars. 
Most UFDs display the by now typical extremely low neutron-capture element abundances \citep[e.g.,][]{Koch13,Frebel14,Ji16a}.
But some UFDs deviate and contain distinctly different chemical signatures: 
Reticulum~II shows the clear signature of a prolific $r$-process event \citep{Ji16b,Ji16c,Roederer16b}; 
and Canes~Venatici~II contains a star that may have an abnormally high [Sr/Ba] ratio \citep{Francois16}.
The diversity of neutron-capture element abundances in UFDs can be interpreted as resulting from highly stochastic production of neutron-capture elements at low [Fe/H] (e.g., \citealt{Lee13}).

The UFD Tucana~II (henceforth Tuc~II) was recently discovered in the Dark Energy Survey \citep{Koposov15a,Bechtol15}. 
It was confirmed to be a galaxy by \citet{Walker16} since it displays a significant velocity
dispersion ($8.6^{+4.4}_{-2.7}$\,{\kms}) and its stars span a range of up to 1\,dex in metallicity.
The low luminosity ($M_V \sim -3.8$) and overall metallicity ($\left<\mbox{[Fe/H]}\right> \sim -2.2$) suggests that Tuc~II stars may contain clues to early nucleosynthesis and the nature of the first stars.

\section{Observations and Abundance Analysis} \label{s:methods}
\begin{deluxetable*}{l|rrrr|rrrr|rrrr|rrrr}
\tablecolumns{17}
\tablewidth{0pt}
\tabletypesize{\footnotesize}
\tabletypesize{\tiny}
\tablecaption{Observed Stars and Abundances\label{tbl:stars}}
\tablehead{\colhead{} & \multicolumn{4}{c}{TucII-006} & \multicolumn{4}{c}{TucII-011} & 
                        \multicolumn{4}{c}{TucII-033} & \multicolumn{4}{c}{TucII-052}}
\startdata
RA (hms) & \multicolumn{4}{c}{22 51 43.06} & \multicolumn{4}{c}{22 51 50.28} & 
     \multicolumn{4}{c}{22 51 08.32} & \multicolumn{4}{c}{22 50 51.63} \\
DEC (dms) & \multicolumn{4}{c}{$-$58 32 33.7} & \multicolumn{4}{c}{$-$58 37 40.2} &
      \multicolumn{4}{c}{$-$58 33 08.1} & \multicolumn{4}{c}{$-$58 34 32.5} \\
$V$\tablenotemark{a} & \multicolumn{4}{c}{18.5} & \multicolumn{4}{c}{17.9} & \multicolumn{4}{c}{18.4} & \multicolumn{4}{c}{18.5} \\
$t_{\rm exp} (min)$\tablenotemark{b} & \multicolumn{4}{c}{55, $2 \times 50$} & \multicolumn{4}{c}{$3 \times 55$, $2 \times 50$} & 
           \multicolumn{4}{c}{$2 \times 50$}    & \multicolumn{4}{c}{$2 \times 50$} \\
S/N\tablenotemark{c} & \multicolumn{4}{c}{10, 20}   & \multicolumn{4}{c}{16, 31} & 
      \multicolumn{4}{c}{17, 31}  & \multicolumn{4}{c}{15, 27} \\
$v_{\rm hel}$ (km/s) & \multicolumn{4}{c}{$-125.3 \pm 0.1$}   & \multicolumn{4}{c}{$-126.1 \pm 0.1$} & 
                  \multicolumn{4}{c}{$-126.8 \pm 0.1$}   & \multicolumn{4}{c}{$-121.4 \pm 0.1$} \\
{\Teff} (K)\tablenotemark{d}         & \multicolumn{2}{c}{$4945 \pm 215$} & \multicolumn{2}{c}{$4900 \pm 200$}  & \multicolumn{2}{c}{$4675 \pm 162$} & \multicolumn{2}{c}{$4670 \pm 150$} & 
                  \multicolumn{2}{c}{$4855 \pm 166$}   & \multicolumn{2}{c}{$4800 \pm 100$}   & \multicolumn{2}{c}{$4900 \pm 256$} & \multicolumn{2}{c}{$4800 \pm 180$}\\
{\logg} (cgs)\tablenotemark{d}         & \multicolumn{2}{c}{$1.90 \pm 0.40$} & \multicolumn{2}{c}{$1.90 \pm 0.40$}  & \multicolumn{2}{c}{$1.00 \pm 0.36$} & \multicolumn{2}{c}{$1.30 \pm 0.20$} &
                  \multicolumn{2}{c}{$1.45 \pm 0.34$}   & \multicolumn{2}{c}{$1.60 \pm 0.20$} & \multicolumn{2}{c}{$1.96 \pm 0.42$} & \multicolumn{2}{c}{$2.10 \pm 0.40$}\\
{\micro} (km/s)\tablenotemark{d}        & \multicolumn{2}{c}{$2.20 \pm 0.26$}   & \multicolumn{2}{c}{$2.40 \pm 0.30$}   & \multicolumn{2}{c}{$1.96 \pm 0.22$} & \multicolumn{2}{c}{$2.20 \pm 0.20$} &
                  \multicolumn{2}{c}{$2.28 \pm 0.23$}   & \multicolumn{2}{c}{$2.20 \pm 0.20$} & \multicolumn{2}{c}{$2.00 \pm 0.30$} & \multicolumn{2}{c}{$2.20 \pm 0.30$} \\
$\mbox{[Fe/H]}$ (dex)\tablenotemark{d}          & \multicolumn{2}{c}{$-3.18 \pm 0.21$} & \multicolumn{2}{c}{$-2.93 \pm 0.14$}  & \multicolumn{2}{c}{$-3.00 \pm 0.19$} & \multicolumn{2}{c}{$-2.78 \pm 0.15$} &
 \multicolumn{2}{c}{$-2.59 \pm 0.22$} & \multicolumn{2}{c}{$-2.52 \pm 0.17$} & \multicolumn{2}{c}{$-3.25 \pm 0.25$} & \multicolumn{2}{c}{$-3.08 \pm 0.16$}
\\
\hline\\
 & $N$ & $\log \epsilon(X)$ & $\sigma$ & [X/Fe] &
   $N$ & $\log \epsilon(X)$ & $\sigma$ & [X/Fe] &
   $N$ & $\log \epsilon(X)$ & $\sigma$ & [X/Fe] &
   $N$ & $\log \epsilon(X)$ & $\sigma$ & [X/Fe]
\\[1mm]
\hline
C\tablenotemark{e}          &   2 &$  5.95$&$   0.22$&$  0.70$&$ 2$&$  5.77$&$   0.30$&$  0.34$&$  2$&$  6.53$&$   0.21$&$  0.70$&$ 2$&$  5.89$&$   0.17$&$   0.71$\\
Na I       &   2 &$  3.21$&$   0.17$&$  0.15$&$ 2$&$  3.87$&$   0.29$&$  0.62$&$  2$&$  3.37$&$   0.04$&$ -0.28$&$ 2$&$  3.15$&$   0.27$&$   0.16$\\
Mg I       &   3 &$  4.80$&$   0.17$&$  0.38$&$ 5$&$  5.33$&$   0.11$&$  0.73$&$  4$&$  4.97$&$   0.34$&$ -0.03$&$ 2$&$  4.80$&$   0.11$&$   0.45$\\
Al I       &   1 &$< 4.77$& \nodata &$< 1.50$&$ 2$&$  2.79$&$   0.65$&$ -0.66$&$  2$&$  2.96$&$   0.73$&$ -0.90$&$ 2$&$  2.75$&$   0.73$&$  -0.45$\\
Si I       &   1 &$< 5.83$& \nodata &$< 1.50$&$ 1$&$  5.09$&$   0.30$&$  0.58$&$  1$&$  5.14$&$   0.23$&$  0.22$&$ 1$&$< 6.26$& \nodata &$ < 2.00$\\
Ca I       &   4 &$  3.48$&$   0.17$&$  0.33$&$ 8$&$  3.89$&$   0.16$&$  0.54$&$  6$&$  4.05$&$   0.25$&$  0.31$&$ 3$&$  3.37$&$   0.13$&$   0.28$\\
Sc II      &   4 &$ -0.13$&$   0.31$&$ -0.10$&$ 5$&$  0.23$&$   0.19$&$  0.08$&$  5$&$  0.14$&$   0.21$&$ -0.42$&$ 5$&$ -0.05$&$   0.31$&$   0.05$\\
Ti II      &   9 &$  1.97$&$   0.23$&$  0.21$&$19$&$  2.02$&$   0.23$&$  0.06$&$ 18$&$  2.23$&$   0.21$&$ -0.13$&$10$&$  2.03$&$   0.16$&$   0.33$\\
Cr I       &   1 &$  1.94$&$   0.33$&$ -0.52$&$ 4$&$  2.32$&$   0.31$&$ -0.32$&$  7$&$  3.03$&$   0.35$&$ -0.02$&$ 5$&$  2.26$&$   0.06$&$  -0.13$\\
Mn I       &   2 &$  1.40$&$   0.72$&$ -0.85$&$ 3$&$  1.41$&$   0.44$&$ -1.02$&$  3$&$  2.14$&$   0.62$&$ -0.70$&$ 3$&$  1.25$&$   0.76$&$  -0.93$\\
Fe I       &  32 &$  4.32$&$   0.21$&$  0.00$&$88$&$  4.50$&$   0.19$&$  0.00$&$101$&$  4.91$&$   0.22$&$  0.00$&$37$&$  4.25$&$   0.25$&$   0.00$\\
Fe II      &   0 & \nodata& \nodata & \nodata&$10$&$  4.47$&$   0.19$&$ -0.04$&$ 15$&$  4.94$&$   0.20$&$  0.03$&$ 2$&$  4.25$&$   0.28$&$   0.00$\\
Co I       &   1 &$< 3.92$& \nodata &$< 2.11$&$ 1$&$  1.98$&$   0.37$&$ -0.01$&$  4$&$  2.41$&$   0.29$&$  0.01$&$ 1$&$< 3.71$& \nodata &$ < 1.97$\\
Ni I       &   1 &$< 4.06$& \nodata &$< 1.02$&$ 1$&$  3.23$&$   0.17$&$  0.01$&$  1$&$  3.60$&$   0.33$&$ -0.03$&$ 1$&$< 3.48$& \nodata &$ < 0.51$\\
Sr II  &   2 &$ -1.51$&$   0.47$&$ -1.20$&$ 2$&$ -2.18$&$   0.43$&$ -2.05$&$  2$&$ -0.62$&$   0.61$&$ -0.90$&$ 2$&$ -1.63$&$   0.50$&$  -1.25$\\
Sr II \tablenotemark{f}  &&$<-0.31$&\nodata&$<0.00$& &$<-0.63$&\nodata&$<-0.50$& &\nodata&\nodata&\nodata& &$<-0.38$&\nodata&$<0.00$\\
Ba II  &   2 &$ -1.85$&$   0.26$&$ -0.85$&$ 1$&$ -2.62$&$   0.30$&$ -1.80$&$  2$&$ -1.56$&$   0.36$&$ -1.15$&$ 2$&$ -2.02$&$   0.29$&$  -0.95$\\
Ba II \tablenotemark{f}  &&$<-1.00$&\nodata&$<0.00$ &&$<-1.82$&\nodata&$<-1.00$& &\nodata&\nodata&\nodata& &$<-1.57$&\nodata&$<-0.5$\\
Eu II      &   1 &$<-0.96$& \nodata &$< 1.70$&$ 1$&$<-1.38$& \nodata &$< 1.10$&$  1$&$<-1.27$& \nodata &$< 0.80$&$ 1$&$<-1.96$& \nodata &$ < 0.77$
\enddata
\tablenotetext{a}{Converted from $g$ and $r$ with formula in \citet{Bechtol15}.}
\tablenotetext{b}{Exposure times for {\TucA} and {\TucB} are listed separately for each night. The seeing was poor the first night.}
\tablenotetext{c}{S/N per pixel (${\sim}0.1{\AA}$) at 5200{\AA} and 6000{\AA}.}
\tablenotetext{d}{LTE (left) and NLTE (right) stellar parameters. LTE uncertainties include systematic errors.}
\tablenotetext{e}{Carbon abundances are already corrected for evolutionary status \citep{Placco14}.}
\tablenotetext{f}{Conservative abundance upper limit, see Figure~\ref{f:spec} and text for details.}
\end{deluxetable*}

We selected four of the brightest high-probability members of Tuc~II from \citet{Walker16}: {\TucA}, {\TucB}, {\TucC}, and {\TucD}.
On 2016 Aug 29-30, we used the Magellan Inamori Kyocera Echelle (MIKE) spectrograph \citep{Bernstein03}
on the Magellan-Clay telescope with a 1\farcs0 slit to obtain spectra
of these stars ($R \sim 22,000$ and $28,000$ on the red and
blue chip, respectively) covering ${\sim}4000-9000$\,{\AA}. 
The seeing was poor on Aug 29 
(1\farcs0$-$3\farcs0) and good on Aug 30 (${\sim}$0\farcs7).
Individual exposures were 50-55 minutes, with 2-5 exposures per star.
The resulting signal-to-noise ratios are modest, though comparable to previous UFD star observations \citep[e.g.,][]{Ji16a}.
Table~\ref{tbl:stars} details our observations.

\begin{figure*}
\begin{center}
  \includegraphics[width=5cm]{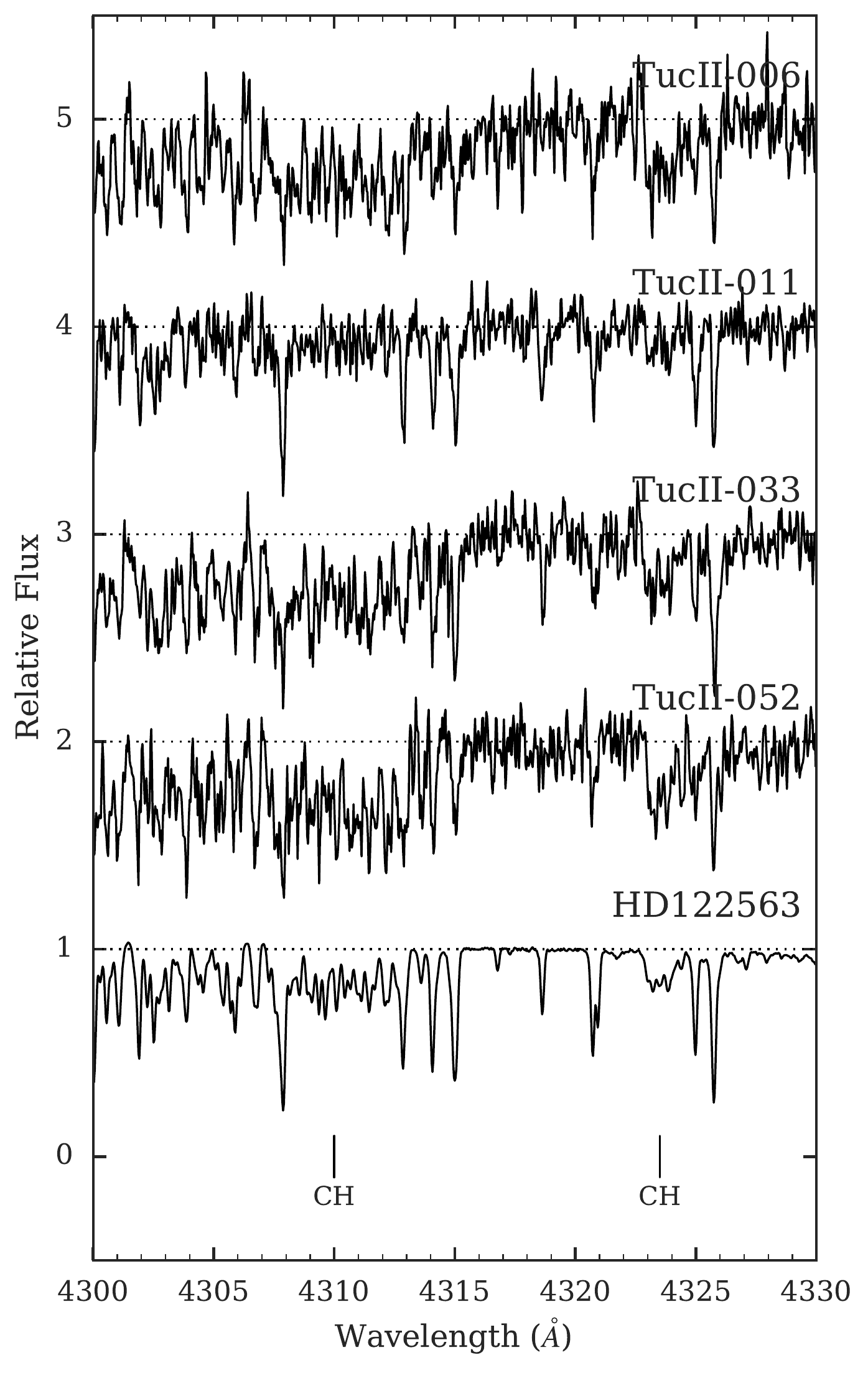}
  \includegraphics[width=4.8cm]{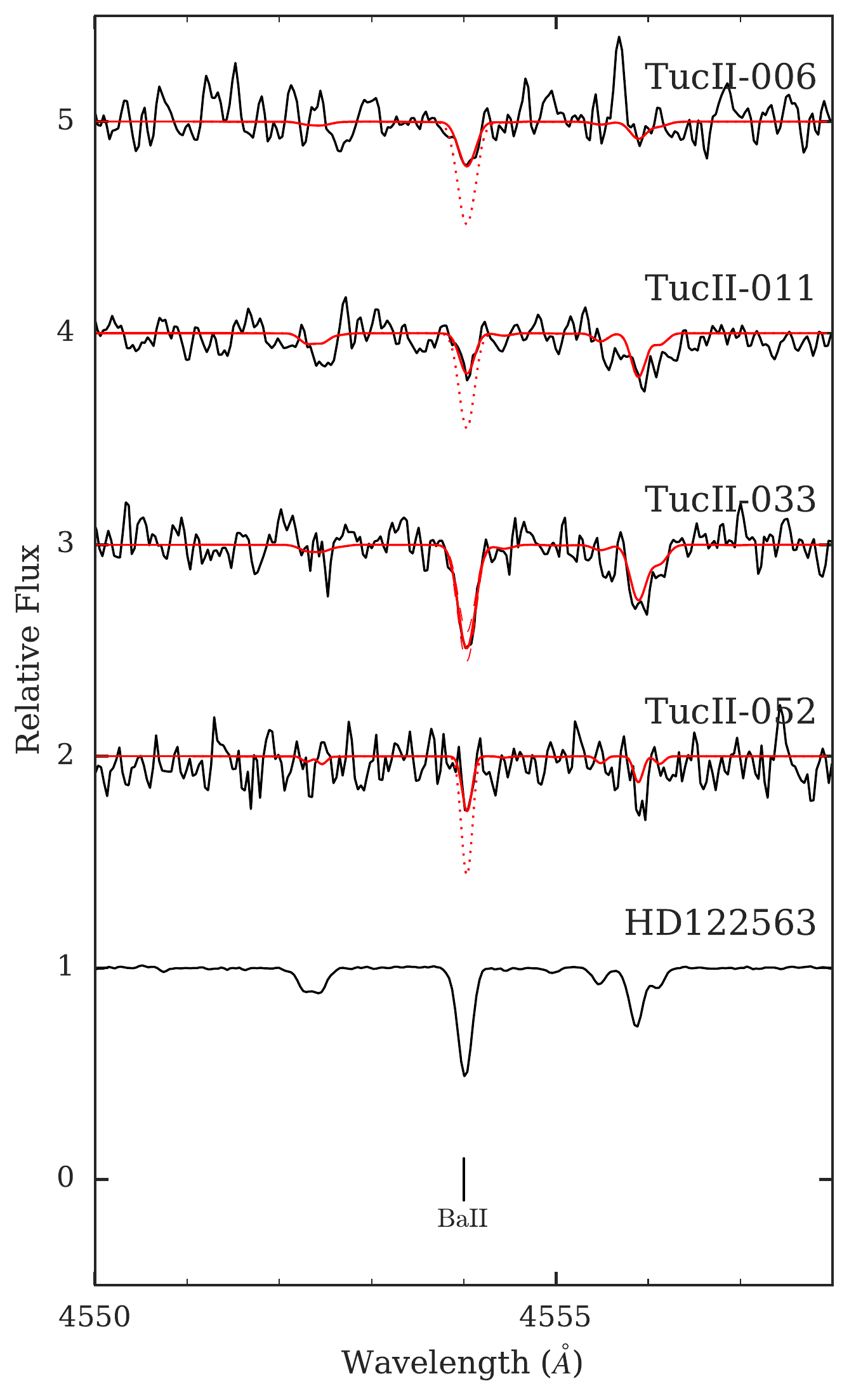}
  \includegraphics[width=5cm]{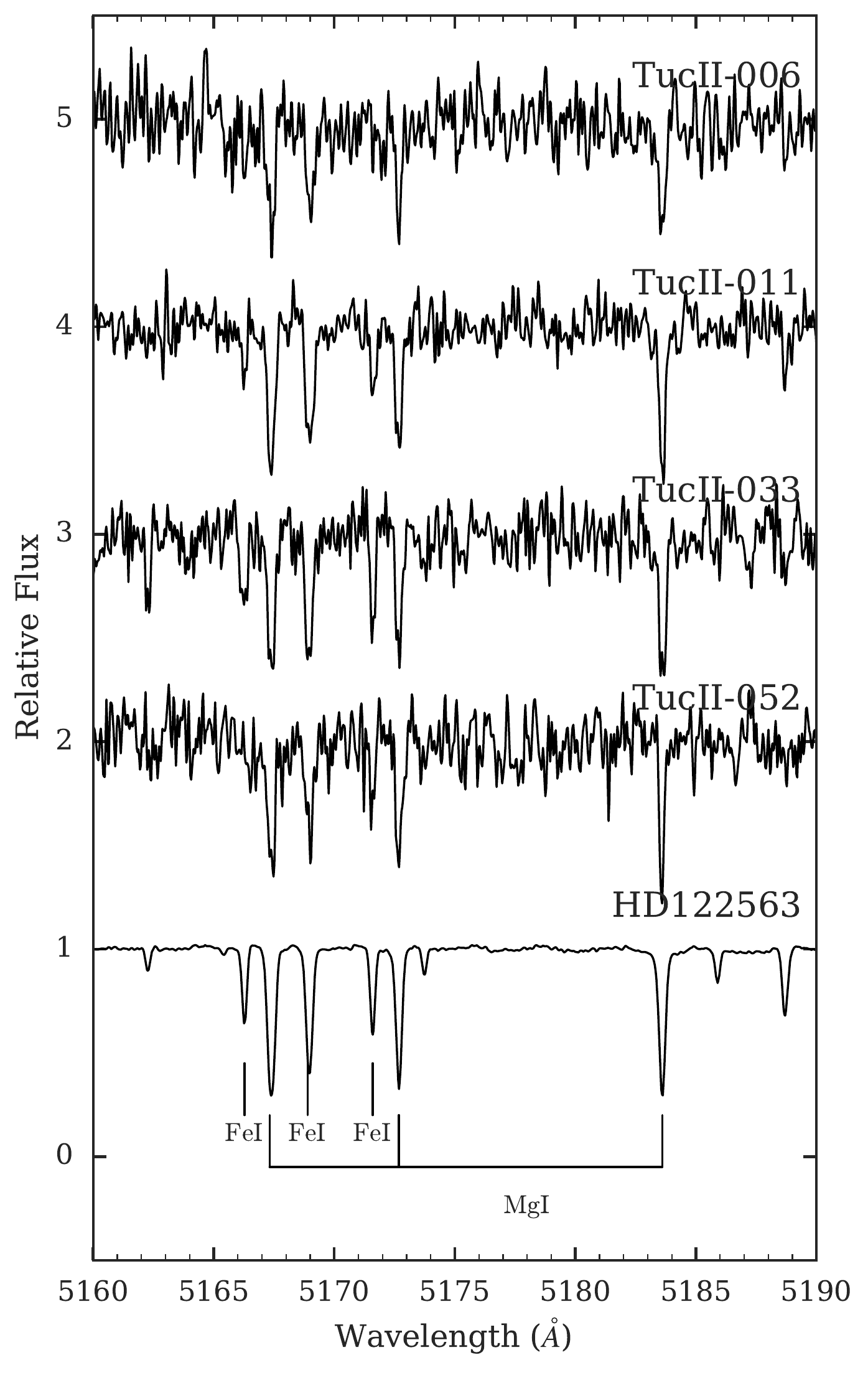}
\end{center}
\caption{Spectral regions around CH band, Ba line, and Mg triplet. HD122563 is shown for comparison.
Around the Ba 4554\,{\AA} line, we show our best-fit synthesis in solid red, and upper limits of [Ba/Fe] = 0, $-1$, and $-0.5$ for {\TucA}, {\TucB}, and {\TucD} in dotted red lines (Table~\ref{tbl:stars}). Dashed red lines around {\TucC} indicate $\pm 0.3$\,dex.
\label{f:spec}}
\end{figure*}

Spectra were reduced with the CarPy MIKE pipeline 
\citep{Kelson03}\footnote{\url{http://code.obs.carnegiescience.edu/mike}}.
We normalized the spectra and determined radial velocity by cross-correlation with 
the Mg triplet near 5200\,{\AA} using the SMH analysis software \citep{Casey14}. 
Heliocentric velocity corrections were determined with \texttt{rvcor}
in \texttt{IRAF}.
Our results match the velocities reported by \citet{Walker16}
within ${\lesssim}2$\,{\kms}, showing no clear evidence for binaries.
Selected spectral regions are shown in Figure~\ref{f:spec}.

We perform a standard abundance analysis of these stars (details given
in \citealt{Frebel13,Ji16c}). The analysis is performed
exactly the same way as in \citet{Ji16c} but we
summarize key points here.
We used SMH to measure equivalent widths and run the MOOG
\texttt{abfind} and \texttt{synth} drivers for stellar parameters and
abundances \citep{Sneden73}.
We use the 2011 MOOG version which accounts for scattering \citep{Sobeck11}.
Stellar parameters are determined spectroscopically through
excitation, ionization, and reduced equivalent width balance \citep[e.g.,][]{Frebel13}.
Stellar parameter uncertainties are determined assuming systematic
uncertainties of 150\,K, 0.3\,dex, and 0.2\,{\kms} for {\Teff},
{\logg}, and {\micro} respectively (see \citealt{Ji16b}).
{\TucA} has no reliable {Fe~II} line detections, so we use its {\Teff} to
determine {\logg} from a $\mbox{[Fe/H]=-3}$, 12\,Gyr isochrone \citep{Kim02}, 
and we adopt a conservative uncertainty of 0.4\,dex.
We use spectral synthesis to determine the abundance of C, Sc, Mn, Sr,
Ba, and some lines of Al and Si. Abundances of other elements are
determined from equivalent width fitting, where the typical uncertainties are
$6{-}12\%$ ($10{-}17\%$ for {\TucA}, which has a noisier spectrum).
Table~\ref{tbl:stars} reports our stellar parameters and abundances.

One-dimensional models invoking the assumption of LTE can produce biased abundances at low [Fe/H]. As such, we also determine NLTE stellar parameters fully spectroscopically, following a new method set out in \citet{Ezzeddine2016}.
{\Teff} values are $\lesssim 100$\,K lower than those derived in LTE, and the three stars with available Fe\,II line measurements have slightly higher {\logg} values ($\leq\,0.3$\,dex). 
All four stars have higher final Fe abundance in NLTE, with differences of $\Delta$[Fe/H] = [Fe/H]$_{\rm NLTE}$ - [Fe/H]$_{\rm LTE}$ on the order of $0.25$\,dex. Increases of this magnitude are expected at these low metallicities and are in accordance with other NLTE-LTE Fe corrections \citep{Lind2012}.
As the majority of literature studies use LTE abundances, our subsequent discussion focuses on the LTE values. However, we consider NLTE iron abundances when discussing the Tuc~II metallicity spread in Section~\ref{s:tuchistory}. Future work should make a concerted effort to address this and other topics with full NLTE-derived abundances.

\citet{Walker16} also determined the stellar parameters and metallicities of these stars from their M2FS data, using a grid of spectra from the Segue Stellar Parameter Pipeline \citep[SSPP,][]{Lee08a}. 
They cover a 50\,{\AA} region around the Mg b triplet. In this region, the MIKE and M2FS spectra have comparable signal-to-noise ratios and spectral resolution. 
Given the limited wavelength range, their stellar parameters are strongly influenced by a prior from photometry \citep{Koposov15a}. 
They determine stellar parameters independently for their repeat Jul 2015 and Sep 2015 spectra.
Our results agree well with at least one of these two stellar parameter determinations.

At first, it appears that our metallicities are significantly lower than those determined by \citet{Walker16}. 
However, they calibrate their model against twilight spectra of the Sun. 
They apply a $-0.3$ offset to {\logg} and a $-0.32$\,dex offset to [Fe/H] to match the solar values.
When we use their stellar parameters and increase our metallicities by $+0.32$\,dex, the [Fe/H] abundances agree to within $<0.15$\,dex for {\TucA}, {\TucC}, and {\TucD}, well within the statistical uncertainties.
However, our metallicities are systematically $\sim0.3$\,dex less for {\TucB}, which is a cooler star.
We verify with the online M2FS spectra\footnote{\url{http://dx.doi.org/10.5281/zenodo.37476
}} that the differences are not due to noise in the data.
This star is somewhat Mg enhanced ($\mbox{[Mg/Fe]} = 0.73$), which may explain the difference as the SSPP grid assumes $\mbox{[Mg/Fe]} = 0.4$.
Otherwise, a difference between the SSPP spectral grid (which was synthesized with \texttt{turbospectrum}) and MOOG for cooler stars may be responsible for the discrepancy.

Given our data quality, we consider most absorption lines below 4000\,{\AA} to
be unreliable, although we have used a few strong and well-detected iron lines. This
restricts our ability to determine abundances of some elements.
In particular, Al abundances are derived from only two lines at 
${\sim}3950$\,{\AA}, and have very large uncertainties.
The Si 3905\,{\AA} line is entirely unreliable, so we only use the 4102\,{\AA} line when available.
Mn abundances should be regarded with caution as they
are derived from the 4030\,{\AA}, 4033\,{\AA}, and 4034\,{\AA} lines.
The Sr 4077\,{\AA} line is clearly detected but has large abundance uncertainties.
We place Eu limits with the 4129\,{\AA} line.

We correct the carbon abundances for the stars' evolutionary status
\citep{Placco14} by assuming $\mbox{[N/Fe]}=0.5$, although the corrections
differ by ${<}0.05$\,dex for $-0.5<\mbox{[N/Fe]}<1.0$. 
In Table~\ref{tbl:stars}, [C/Fe] has been corrected by $+0.05$\,dex ({\TucA}), $+0.74$\,dex ({\TucB}), $+0.47$\,dex ({\TucC}), and $+0.01$\,dex ({\TucD}).
Even after applying the large correction, {\TucB} is not carbon enhanced.
The other three stars are just past the threshold of the Carbon-Enhanced Metal-Poor (CEMP) definition ($\mbox{[C/Fe]}\geq 0.7$, \citealt{Aoki07}).

The lines of neutron-capture elements Sr and Ba are detected in all stars, although in some stars the line depths are only somewhat larger than the noise level.
In Table~\ref{tbl:stars}, we list the abundances measured from all detected features as well as conservative upper limits (see middle panel of Figure~\ref{f:spec} for an example). 
All four stars clearly have $\mbox{[Sr,Ba/Fe]}<0$, making three of them CEMP-no stars.

\begin{figure*}
\begin{center}
  \includegraphics[width=18cm]{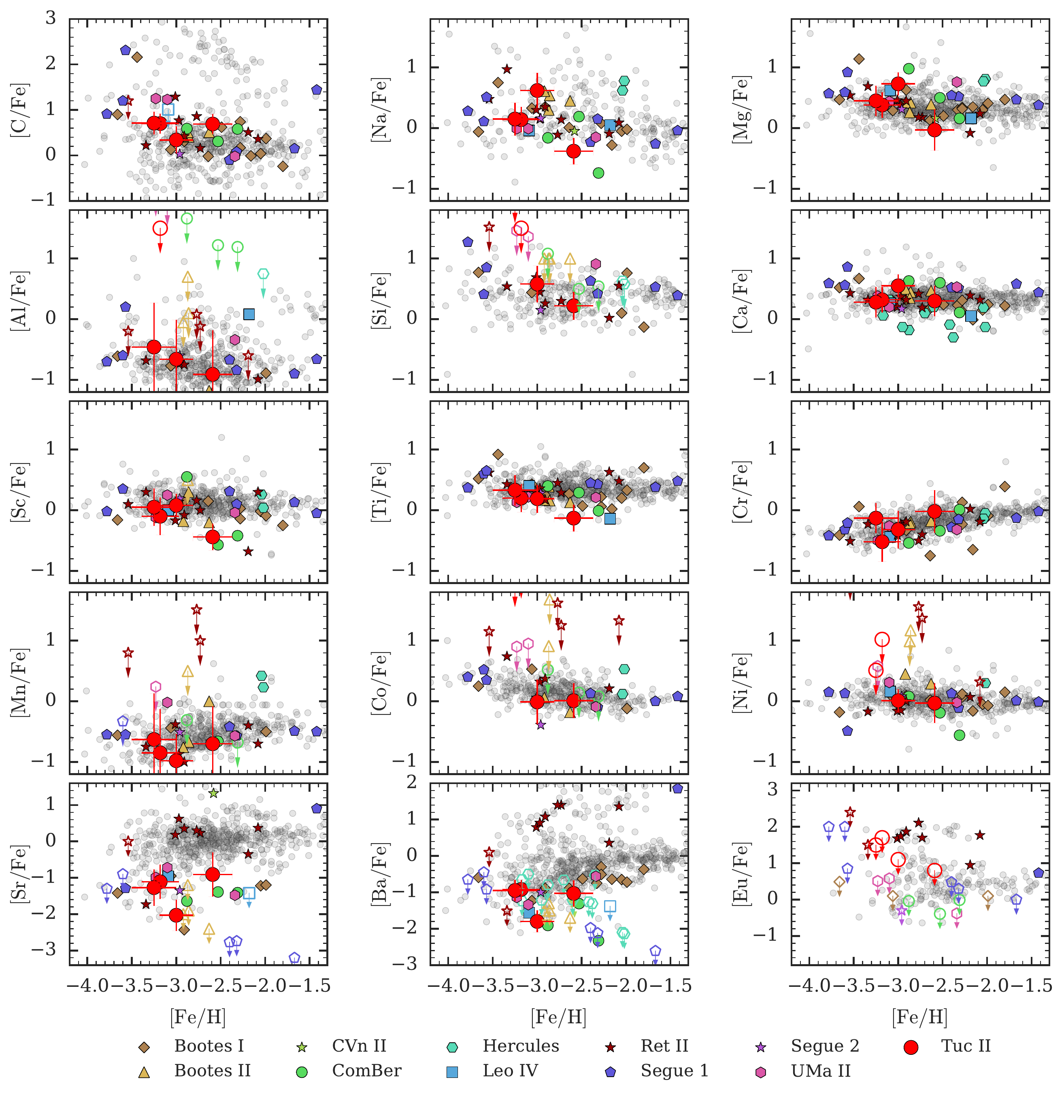}
\end{center}
\caption{Abundances of elements in Tuc~II (large red points, this
  work), 
  other UFDs (colored points, see references in \citealt{Ji16c},
  Sr and Ba for one star from \citealt{Roederer16b}), 
  and halo stars \citep[gray points,][]{Frebel10}. 
  Open points with arrows indicate upper limits. 
  From low to high [Fe/H], the Tuc~II stars are {\TucD}, {\TucA},
  {\TucB}, and {\TucC}.
  {\TucD} and {\TucA} have similar overall abundances corresponding to
  Group~II CEMP-no stars. {\TucB} is not carbon-enhanced and appears to have 
  different Na, Mg, Ca, Sr, and Ba than the two CEMP-no stars.
  The higher metallicity star {\TucC} shows evidence for extended
  chemical enrichment in Tuc~II.
  \label{f:abunds}}
\end{figure*}

Figure~\ref{f:abunds} shows the abundances of our four Tuc~II stars compared to those of equivalent halo stars and stars in other UFDs.
Overall, the Tuc~II stars have similar abundances as other UFD stars, with typical halo-like abundances of elements up to Ni, and low neutron-capture element abundances.
Nevertheless, there are several interesting abundance differences between these four Tuc~II stars that we now consider.

\section{Pop~III signatures in Tucana~II} \label{s:popiii}
We first focus on the three extremely metal-poor (EMP) stars {\TucA}, {\TucB}, and {\TucD}, with $\mbox{[Fe/H]}\leq-3$. These are the stars more likely to trace unique Pop~III nucleosynthesis signatures.

{\TucA} and {\TucD} are CEMP-no stars ($\mbox{[C/Fe]}=0.7$, $\mbox{[Sr,Ba/Fe]}<0$).
They are only just past the CEMP threshold, but the C, Fe, Na, and Mg abundances place these stars squarely as Group~II CEMP-no stars, according to the classification of \citealt{Yoon16}.
All elemental abundances of {\TucA} and {\TucD} are consistent with being identical given the uncertainties, as might be expected if both stars formed from the same star cluster \citep[e.g.,][]{BlandHaw10}. 
The high abundance precision required to test the cluster hypothesis likely requires much higher S/N data that could be obtained with 30m class telescopes (e.g., G-CLEF on the Giant Magellan Telescope, \citealt{GCLEF14}).

{\TucB} is an extremely metal-poor star ($\mbox{[Fe/H]}=-3$) but is not carbon enhanced ($\mbox{[C/Fe]}=+0.34$ after positive correction).
It is just the third EMP star known in a UFD that is not carbon enhanced (of 18 total EMP stars in 11 UFDs).
The other two non-CEMP stars in UFDs are DES\,J033531$-$540148 in Ret~II \citep{Roederer16b,Ji16c} and Boo-980 in Bootes~I \citep{Frebel16}.
This CEMP fraction ($83\%$) is somewhat higher than the halo \citep{Placco14} but consistent with expectations for UFDs \citep{Salvadori15}.
Besides carbon, {\TucB} differs from the Tuc~II CEMP-no stars in having especially low neutron-capture element abundances ($\mbox{[Sr,Ba/Fe]}\sim-2$) as well as somewhat enhanced $\mbox{[Na/Fe]}\sim0.6$, $\mbox{[Mg/Fe]} \sim 0.7$, and $\mbox{[Ca/Fe]}\sim0.55$.
The abundances of {\TucB} are thus qualitatively different from those of {\TucA} and {\TucD}, likely requiring at least two different types of metal sources as explanation.
As all three stars have $\mbox{[Fe/H]} \lesssim -3$, this could suggest 
that Pop~III stars produce at least two distinct types of yields (e.g., \citealt{Cooke14, Ji15}; and in contrast to, e.g., \citealt{Salvadori15}).
An interesting alternate scenario is if different metals created from a single source were to mix differently into the surrounding gas \citep{Sluder16}.
We note high Na and Mg are also found in DES\,J033531$-$540148 (a non-$r$-process star in Reticulum~II), but not in Boo-980.

[Sr/Ba] can in principle provide insight into the origin of the neutron-capture elements in Tuc~II.
We caution against over-interpreting this ratio for our stars, as Sr and Ba have significant abundance uncertainties. But at face value, the three EMP stars all appear to have Sr and Ba detections with $\mbox{[Sr/Ba]}\sim-0.3$.
Empirically from metal-poor halo stars, the $r$-process produces $\mbox{[Sr/Ba]} \sim -0.3$ and the metal-poor $s$-process produces $\mbox{[Sr/Ba]} \lesssim -1$ (computed from $r$-II and CEMP-$s$ stars in \citealt{Frebel10}).
The [Sr/Ba] ratios of stars in Tuc~II thus appear to disfavor the $s$-process as the source of these elements.
However, the lowest metallicity spinstar models ($15-40\,M_\odot$, $Z \sim 10^{-5}Z_\odot$) can also produce $\mbox{[Sr/Ba]}\sim -0.5$ \citep{Frischknecht16}.

In a UFD, it is possible to place loose constraints on the neutron-capture element yields, as the galactic environment restricts gas dilution masses to the range $M_{\rm H} \sim 10^{6\pm1}M_\odot$ \citep{Ji16b}.
For core-collapse supernova models producing Sr in neutrino-driven winds \citep[e.g.,][]{Wanajo13}, the overall yield of $M_{\rm Sr} \sim 10^{-6} M_\odot$ results in $\mbox{[Sr/H]} \sim -5 \pm 1$, consistent with the [Sr/H] ratios observed in Tuc~II\footnote{In these models, Ba is produced with $\mbox{[Sr/Ba]} \sim 0-0.4$ if the proto-neutron star has mass ${\geq}2\,M_\odot$}.
In contrast, a single $15-40\,M_\odot$ $Z \sim 10^{-5}Z_\odot$ spinstar produces $M_{\rm Sr} \sim 10^{-8 \pm 1} M_\odot$ \citep{Frischknecht16}. This would result in $\mbox{[Sr/H]} \sim -7 \pm 2$, lower than what is found in our Tuc~II stars.

\citet{Yoon16} hypothesize the Group~II CEMP-no stars (i.e., {\TucA}, {\TucD}) formed out of gas enriched only by faint, low-energy Pop\,III supernovae \citep[e.g.,][]{Heger10,Nomoto13,Cooke14}.
However, if faint supernovae must be invoked to produce enhanced [C/Fe] seen in some Tuc~II stars, it seems unlikely that any neutron-capture material produced deep in the core of the massive star will be able to escape.
The neutron-capture elements found in the CEMP-no stars would then have to be synthesized in other ways, either elsewhere in the star or from a completely different source. 
A possible alternative is if core material escapes through jets, as some jet supernovae may also produce carbon enhanced metal yields \citep{Tominaga07}.
We note that the apparent ubiquity of neutron-capture elements in metal-poor stars \citep{Roederer13} suggests that there should be a mechanism capable of producing these elements early on, even if only in very small amounts.

\section{Extended star formation in Tucana~II} \label{s:tuchistory}
One of our four stars ({\TucC}) is relatively metal-rich ($\mbox{[Fe/H]}=-2.6$).
Because of the higher Fe content, simple homogeneous chemical evolution models would imply that this star formed later than the other three stars. Inhomogeneous metal mixing is an alternate 
possibility to produce this star  \citep{Frebel12,Karlsson13,Webster15}.
This star has similar [C, Sr, Ba/Fe] ratios to the CEMP-no stars, but much lower [Na, $\alpha$, Sc/Fe] $\lesssim 0$.
The abundance uncertainties are large, but if these differences
are all true then one explanation is that this star has formed from 
gas additionally enriched both by Type~Ia supernovae (decreasing [X/Fe] for most elements) 
and from AGB stars (increasing [C,Sr,Ba/Fe]).
{\TucC} thus provides evidence for more extended chemical enrichment in Tuc~II, in contrast
to the smallest UFDs like Segue~1 ($M_V=-3.8$ and $-1.5$ for Tuc~II and Segue~1, respectively, \citealt{Koposov15a,Simon11}).
More detailed investigations into the galaxy formation history require either very accurate photometry \citep[e.g.,][]{Brown14} or a much larger sample of stellar metallicities \citep[e.g.,][]{Kirby11b}.
According to the one-shot enrichment criteria in \citet{Frebel12}, we thus do not consider Tuc~II to be a ``first galaxy'' candidate.
Segue~1 thus remains the only known galaxy to date unambiguously satisfying the first galaxy criteria \citep{Frebel14}.

The metallicity distribution function (MDF) of dwarf galaxies can
provide additional insight into their formation history \citep[e.g.,][]{Kirby11b,Webster15}.
Unfortunately, \citet{Walker16} have too few stars to formally resolve the 
metallicity dispersion $\sigma_{\mbox{[Fe/H]}}$ of Tuc~II, but the metallicity range of their 
probable members is ${\sim}1$\,dex.
When interpreting MDFs, one possible concern is that metallicities derived from LTE may
be systematically offset due to NLTE effects.
Based on our NLTE stellar parameters, the average metallicity increases by ${\sim}0.2$\,dex. 
The correction is larger in more metal-poor stars, so the overall metallicity range could 
shrink by ${\sim}0.1$\,dex.
Metallicity shifts can in principle affect the interpretations of chemical evolution models fit to MDFs \citep[e.g.,][]{Kirby11b}.
However, we note that a $0.2$\,dex increase in metallicity could also be compensated by a $60\%$ increase in supernova iron yields.

\citet{Walker16} point out that the orbit of Tuc~II is consistent with it being a member of the LMC system.
This raises the question of whether the LMC environment might somehow affect the formation history of this galaxy.
Indeed, hierarchical structure formation simulations suggest that most present-day Milky Way subhalos associated with UFDs fell into the Milky Way as members of larger systems \citep{Wetzel15}.
However, UFDs complete ${>}80\%$ of their star formation by $z=6$ \citep{Brown14} and do not tend to accrete into larger systems until well after $z=6$ \citep{Wetzel15}. 
Consequently, their star formation history is probably more influenced by reionization than by environmental effects.
Furthermore, a galaxy of Tuc~II's luminosity is unlikely to have more than one star-forming progenitor halo (Griffen et al., in prep).
Given their low mass and mostly isolated formation histories, high-resolution hydrodynamic zoom-in simulations of UFDs should be relatively inexpensive. 
We suggest that statistical samples of UFD simulations could be a fruitful path to understanding questions such as inhomogeneous metal mixing and the impact of different reionization models on the formation history of these galaxies.

\acknowledgements
We thank Ani Chiti for helpful discussions. APJ and AF are supported by NSF-CAREER grant AST-1255160. AF acknowledges support from the Silverman (1968) Family Career Development Professorship. This work benefited from support by the National Science Foundation
under Grant No. PHY-1430152 (JINA Center for the Evolution of the
Elements).
A.~R.~C. was supported by the European Union FP7 programme through ERC grant number 320360.
This work made extensive use of NASA's Astrophysics Data System
Bibliographic Services and the python libraries 
\texttt{numpy},
\texttt{scipy},
\texttt{matplotlib},
\texttt{pandas},
\texttt{seaborn},
and \texttt{astropy} \citep{astropy}.

\end{document}